\documentclass[twocolumn,showpacs,preprintnumbers,amsmath,amssymb,prl,superscriptaddress]{revtex4}
\usepackage{graphicx}

\begin{document}

\newcommand {\be}{\begin{equation}}
\newcommand {\ee}{\end{equation}}
\newcommand {\bea}{\begin{eqnarray}}
\newcommand {\eea}{\end{eqnarray}}
\newcommand {\nn}{\nonumber}

\title{Universal scaling at field-induced magnetic phase transitions} 

\author{Omid Nohadani}
\affiliation{Department of Physics and Astronomy, University of Southern
California, Los Angeles, CA 90089-0484}
\email{O. Nohadani: nohadani@usc.edu}

\author{Stefan Wessel}
\affiliation{Institut f\"ur Theoretische Physik, ETH-H\"onggerberg, 
CH-8093 Z\"urich, Switzerland}

\author{B. Normand} 
\affiliation{D\'epartement de Physique, Universit\'e de Fribourg, 
CH-1700 Fribourg, Switzerland} 

\author{Stephan Haas}
\affiliation{Department of Physics and Astronomy, University of Southern
California, Los Angeles, CA 90089-0484}

\pacs{PACS numbers: }

\begin{abstract}

We study field-induced magnetic order in cubic lattices of dimers 
with antiferromagnetic Heisenberg interactions. The thermal critical 
exponents at the quantum phase transition from a spin liquid to a 
magnetically ordered phase are determined from Stochastic Series 
Expansion Quantum Monte Carlo simulations. These exponents are
independent of the interdimer coupling ratios, and converge to the 
value obtained by considering the transition as a Bose-Einstein 
condensation of magnons, $\alpha_{\rm BEC} = 1.5$. The scaling 
results are of direct relevance to the spin-dimer systems TlCuCl$_3$ 
and KCuCl$_3$, and explain the broad range of exponents reported for 
field-induced ordering transitions.

\end{abstract}

 
\maketitle

A quantum spin liquid is characterized by a finite magnetic correlation 
length, which is inversely proportional to the energy gap, $\Delta_0$, 
between the singlet ground state and the lowest triplet excitation. 
Application of an external magnetic field causes a linear reduction of 
$\Delta_0$ by the Zeeman effect. At zero temperature there is a 
field-induced quantum phase transition (QPT), at a lower critical field 
$h_{c1} \equiv g \mu_{\rm B} H_{c1} = \Delta_0$, from the spin liquid 
to a magnetically ordered state. In this phase the uniform magnetic 
polarization then rises continuously from zero to saturation at a 
second critical field $h_{c2}$. Such transitions have been observed 
in a variety of quantum antiferromagnets, including 
$\rm Cu_2(C_5H_{12}N_2)_2Cl_4$\cite{cuhpcl}, 
$\rm KCuCl_3$\cite{rsttktmg,kcucl}, 
and $\rm TlCuCl_3$\cite{rsttktmg,tlcucl,rtokuokh}, which consist of weakly 
coupled, low-dimensional structural units such as spin ladders or dimers. 
In the ordered phase, the uniform magnetization is accompanied by a 
long-ranged antiferromagnetic (AF) spin ordering in the plane 
perpendicular to the field, which persists up to a finite transition 
temperature, $T_c$, determined by the smallest coupling constant between 
the magnetic structural units.

The compounds $\rm KCuCl_3$ and $\rm TlCuCl_3$ provide an excellent 
illustration of the field-induced QPT from a dimer spin liquid to a 
three-dimensional (3D) ordered phase. These systems consist of dimer 
units formed by pairs of $S = 1/2$ $\rm Cu^{2+}$ ions 
with significant interdimer superexchange interactions in all three 
spatial directions \cite{kcucl,tlcucl}. Because these interactions are 
considerably weaker in KCuCl$_3$ than in TlCuCl$_3$, the magnon modes 
of the former have a narrow band and a large spin gap ($\Delta_0$ = 
2.6meV), while the latter represents a wide-band system with a small 
gap ($\Delta_0$ = 0.7meV). The wave vector ${\bf Q}$ of the 3D AF order 
is the location in momentum space of the band minimum, $\Delta_0$, of 
the triplet magnon excitations, which are gapped in the disordered 
phase. The field-induced QPT may be regarded as a condensation of 
magnons in the lowest triplet branch, which becomes the gapless 
excitation, or Goldstone mode, corresponding to rotations of the 
staggered magnetic moment about the field axis and reflecting the 
broken symmetry of the system \cite{matsumoto}. The continuous evolution 
of the ordered moments is then equivalent to a field-dependent density 
of condensed magnons, and the magnetic field represents an 
experimentally accessible tuning parameter for the QPT. 

The suggestion that this transition may be described as the 
Bose-Einstein condensation (BEC) of hard-core bosons \cite{bec} 
has motivated extensive experimental and theoretical studies of the 
critical properties of such systems. Particular attention is given to 
the scaling exponent $\alpha$ which determines the power-law dependence 
\begin{equation} 
\label{critical_exponent}
T_c (h) \propto |h - h_{c}|^{\frac{1}{\alpha}}
\end{equation}
of the ordering temperature on the field $h$ in the vicinity of the QPTs 
at $h_c \equiv h_{c1}$ and $h_{c2}$. Different experimental studies 
indicate that the precise value of this exponent at $h_{c1}$ depends on 
the material, as well as on the measuring technique. The exponents $\alpha 
\approx 1.8 - 2.3$ extracted from experimental studies of KCuCl$_3$ and 
TlCuCl$_3$ \cite{rsttktmg,kcucl,tlcucl,rtokuokh} are all larger than the 
value $\alpha_{\rm BEC} = 3/2$ obtained from the BEC scenario for 3D 
systems in which the magnon modes have quadratic dispersion at the band 
minimum \cite{bec}. The analytic bond-operator approach, applied with the 
parameters of KCuCl$_3$ and TlCuCl$_3$, also yields the respective values 
2.6 and 1.8 \cite{rmpc}, while recent Quantum Monte Carlo simulations for 
weakly coupled dimers \cite{wessel} suggest that $\alpha \sim 2.3$. In both 
experiments and simulations, precise extraction of the critical exponent is 
complicated by the fact that typically only rather few data points are 
available close to the QPT. In addition, the exponent may not reflect 
universal critical behavior if the temperature exceeds an (unknown) 
upper bound for the validity of the BEC approach. However, it is also 
possible that the scaling behavior is simply not universal, in that the 
exponent depends on certain system parameters, most notably the strength 
of the interdimer interactions. 

\begin{figure}[t!]
\includegraphics[width=8.5cm]{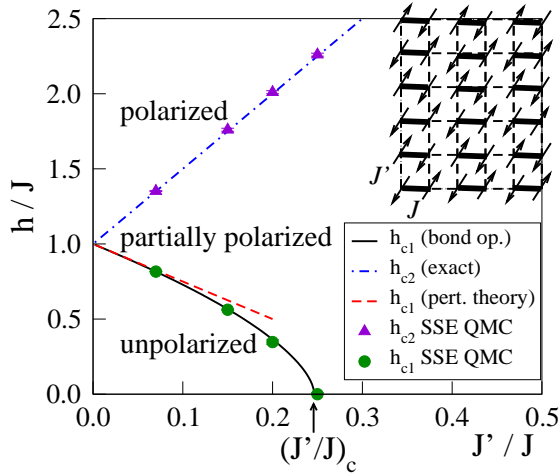}
\vspace{-2cm}
\caption{\label{phase_diagram} 
Phase diagram of 3D AF coupled dimers in a magnetic field. The solid 
line represents $h_{c1}$ and the dot-dashed line $h_{c2}$. Circles 
and triangles are zero-temperature fits obtained from SSE QMC results. 
Inset: one layer of coupled dimers. }
\end{figure}

To analyze the scaling behavior associated with field-induced magnetic 
order, we study a Heisenberg model of coupled spin-1/2 dimers with the 
Hamiltonian
\bea
H = \sum_{\langle i,j \rangle} J_{ij} {\bf S}_i \cdot {\bf S }_j
- h \sum_i S^z_i .
\eea
The AF coupling constants $J$ and $J'$ between nearest-neighbor sites 
$i$ and $j$ denote respectively the intra- and interdimer superexchange 
interactions for the cubic geometry of which a planar section is 
illustrated in the inset of Fig.~\ref{phase_diagram}. Because the 
K/TlCuCl$_3$ structure has no magnetic frustration, the AF cubic dimer 
system is expected to illustrate the same generic similarities and 
differences between large- and small-gap systems as those obtained in 
the experimental studies to date.

We take as the order parameter the component of staggered 
magnetization perpendicular to the applied field,
\begin{equation}
\label{perp_stagg_mag}
m_{\rm s}^{\perp} = \sqrt{\frac{S^\perp_{\rm s}}{L^3}},
\end{equation}
where $L$ denotes the linear system size and $S^\perp_{\rm s}$ the 
perpendicular staggered structure factor, defined as
\begin{equation}
\label{stagg_stracture_factor}
S^\perp_{\rm s} = \frac{1}{L^3}\sum_{\langle i,j \rangle} (-1)^{i+j}
 \langle S^{x}_i S^{x}_j \rangle .
\end{equation}
This quantity is calculated using the Stochastic Series Expansion Quantum 
Monte Carlo (SSE QMC) method of Refs.~\cite{sse} and \cite{sandvik_green}. 
The staggered structure factor is sampled ``on the run" as loop 
updates are constructed \cite{troyer}, with implementation of the 
directed-loop algorithm \cite{directed_loop} to minimize bounce 
probabilities within a loop. This procedure yields a significantly 
higher sampling efficiency than the conventional operator-loop 
update: close to criticality ($J' \ge 0.2 J$ and $h \le 0.7 J$), 
where the maximum number of runs is required, autocorrelation times 
are reduced by factor of up to 8. Simulations were performed for 
cubic clusters of up to 8$\;\!$000 spins, and for temperatures down 
to $T = 0.03J$, except at very small $J'$ where smaller values are 
readily obtainable.

The zero-temperature phase diagram of the coupled dimer system is 
shown in Fig.~\ref{phase_diagram}. In the absence of a magnetic field 
the computed quantum critical point, separating the spin-disordered, 
dimer valence bond solid (DVBS) phase from the ordered AF phase, occurs 
at $(J'/J)_c = 0.255(5)$. Both phases are unpolarized at $h = 0$. 
Their elementary magnetic excitations have the same fundamental 
difference as for the field-induced QPT,
except that the coupling-induced ordered phase 
has two gapless Goldstone modes (associated with the 
broken $O(3)$ symmetry) corresponding to the phase fluctuations of the 
staggered moment. The uniform magnetization varies continuously with 
applied field from zero in the DVBS phase ($h \le h_{c1}$) to full 
polarization at $h \ge h_{c2}$. In the ordered phase the staggered 
moment first increases with $h$ beyond $h_{c1}$ (also for $J' > 
J_c^\prime$) before falling continuously to zero at $h_{c2}$.

\begin{figure}[t!]
\includegraphics[width=7.5cm,height=7cm]{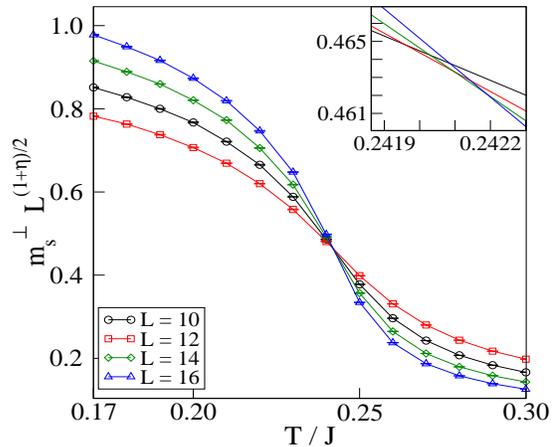}
\vspace{-0.5cm}
\caption{\label{s_mag} Scaled staggered magnetization at constant 
field $h = 0.8 J$ and interdimer coupling $J' = 0.25 J$, for a range 
of system sizes. $T_c$ is determined from the intersection of the four 
curves (main panel), and its error from the extremal intersection points 
(inset). }
\end{figure}

The values of $h_{c2}$ obtained from SSE QMC simulations (triangles 
in Fig.~\ref{phase_diagram}) coincide with the exact result, which is 
classical because quantum fluctuations are entirely suppressed
in the saturated regime. By contrast, quantum corrections are essential 
to the determation of $h_{c1}$, where a 
straightforward perturbative expansion (dashed line) departs from the 
numerical data (circles) for $J' > 0.08J$. However, the self-consistent 
bond-operator solution (solid line) provides an accurate description of 
the numerical results at all interdimer couplings, including those close 
to $J_c^\prime$. The quantitative success of this theory for KCuCl$_3$ 
and TlCuCl$_3$ may be ascribed to the 3D nature of the ordered phase. 
\cite{matsumoto}

Turning to finite-temperature properties, the transition temperatures
for the ordered phase are determined by finite-size scaling of the order 
parameter $m_{\rm s}^{\perp}$. Fig.~\ref{s_mag} shows one example of this 
procedure: in 3D and at the critical temperature, $m_{\rm s}^{\perp}$ at 
fixed field is expected to obey the scaling relation $m_{\rm s}^{\perp} 
\propto L^{(1-z+\eta )/2}$, where the dynamical exponent $z$ is zero
for a thermal transition. \cite{troyer_exponent} For a system with  
spontaneous breaking of the residual $O(2)$ symmetry the correlation exponent 
can be taken from the classical 3D $XY$ model, for which $\eta = 0.0381(2)$ 
\cite{eta}. In the vicinity of the critical 
temperature $T_c$, the scaled $m_{\rm s}^{\perp}$ is linear in $T$ 
\cite{sandvik_critical}, and the intersection of the curves obtained 
for different system sizes $L$ determines $T_c$ at all fields $h \in 
[h_{c1},h_{c2}]$ for a given interdimer coupling $J'$. Magnification 
of the intersection region (inset, Fig.~\ref{s_mag}) reveals a set of 
intercepts within a temperature window $\Delta T \sim 0.0005J$, which 
determines $T_c(h)$ to very high accuracy. \cite{runs}  

\begin{figure}[t!]
\includegraphics[width=8.5cm,height=7cm]{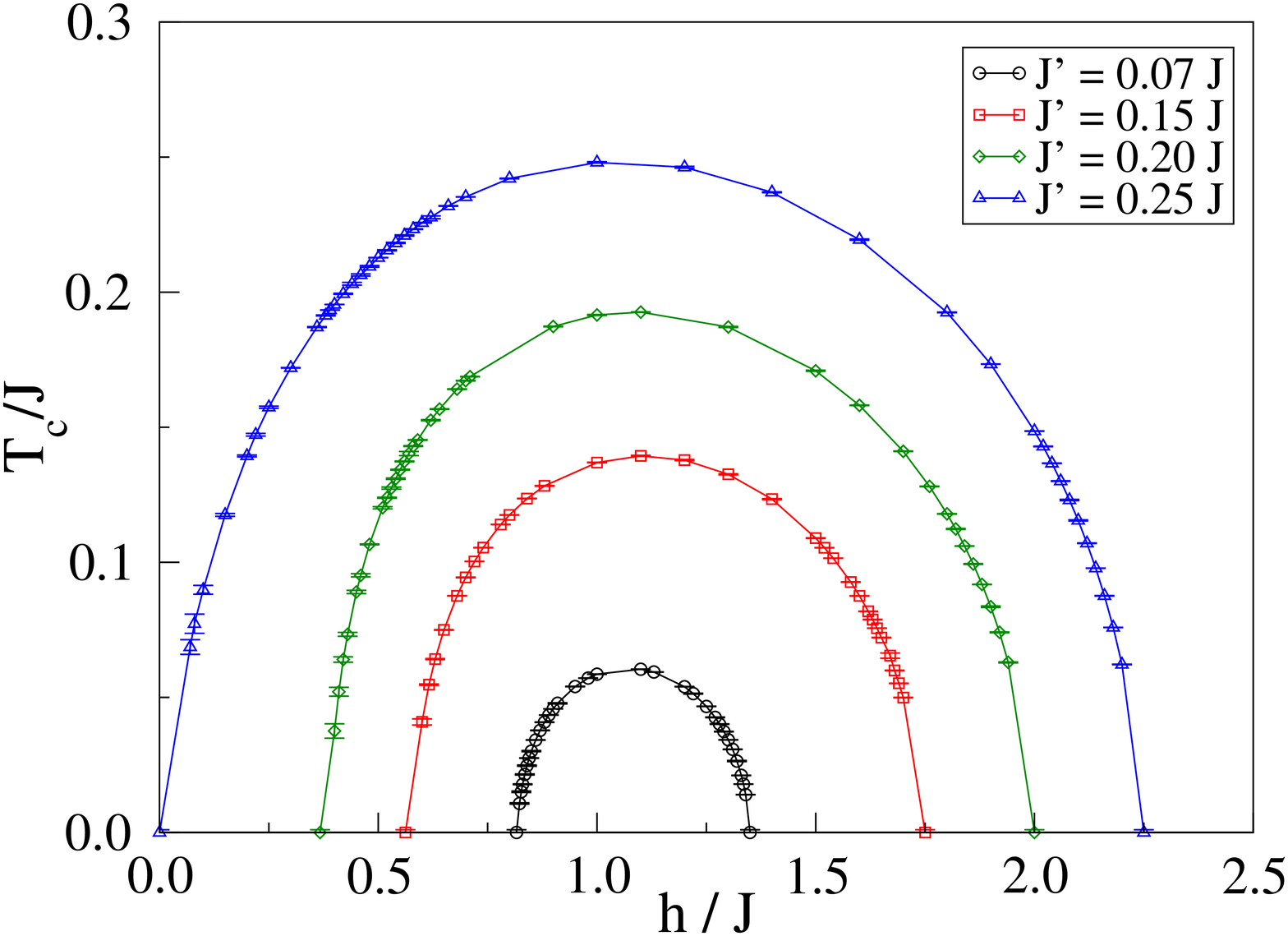}
\vspace{-0.5cm}
\caption{\label{full_phase_diagram} Critical temperature for field-induced
order in 3D systems of coupled dimers for a range of values of $J'$. 
Symbols are taken directly from finite-size scaling of SSE QMC simulations, 
except for the $T = 0$ points which are extracted from fits of $T_c (h)$. Solid 
lines are guides to the eye.}
\end{figure}

Fig.~\ref{full_phase_diagram} shows the critical temperatures $T_c(h)$ 
obtained for four interdimer coupling ratios, $(J'/J)$ = 0.07, 0.15, 0.20, 
and 0.25 \cite{runs}, which span the range from narrow-band, large-gap 
systems to the wide-band, small-gap paradigm. Each coupling ratio 
corresponds to a vertical section in Fig.~\ref{phase_diagram}, where 
the values of $h_{c1}$ and $h_{c2}$ in Fig.~\ref{full_phase_diagram} mark 
the boundaries of the phase with finite induced staggered order; $h_{c1}$ 
vanishes for $J' > J_c^\prime$. The weak asymmetry of the $T_c(h)$ curve 
about the midpoint field, $(h_{c1} + h_{c2})/2$, becomes more pronounced 
for stronger couplings due to changes at the extrema of the magnon 
dispersion relations determining the transitions \cite{matsumoto}. The 
values of $T_c (h)$ obtained by the systematic scaling procedure confirm, 
within the error bars, the location of the dips in uniform magnetization 
reported in previous numerical studies \cite{wessel}, and used to 
construct the experimental phase diagram of $\rm TlCuCl_3$ \cite{tlcucl}. 

We now focus on the scaling properties of $T_c(h)$ close to $h_{c1}$ and 
$h_{c2}$. The scaling exponent $\alpha$ in Eq.~(1) is obtained by fitting 
the data shown in Fig.~\ref{full_phase_diagram} with $h_c$ and $\alpha$ 
as free parameters. We fit this data within a varying window, 
{\it i.e.}~for given $J'/J$ we define
\begin{equation}
x_1(h) = \frac{h - h_{c1}}{h_{c2} - h_{c1}}, \ \ \ \ \ \
x_2(h) = \frac{h_{c2} - h}{h_{c2} - h_{c1}},
\end{equation}
as normalized measures of the proximity to the lower and upper critical 
fields. Using a minimum of 6 data points in a given window $x_1$ or $x_2$, 
the exponents $\alpha_1$ and $\alpha_2$ are then determined by non-linear 
curve fitting. The fitted exponents are crucially dependent both on
the number of points and on the exact critical field 
(Fig.~\ref{phase_diagram}). \cite{errors}

\begin{figure}[t!]
\includegraphics[width=7.5cm]{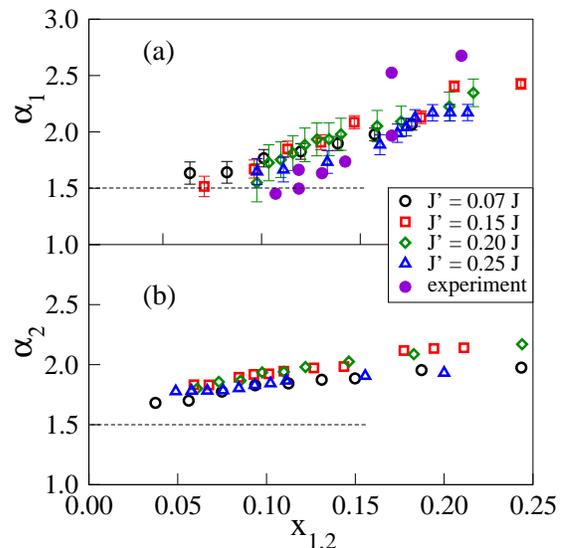}
\vspace{-0.4cm}
\caption{\label{alphas} Critical exponents $\alpha_1$ at $h_{c1}$ (a) and 
$\alpha_2$ at $h_{c2}$ (b), shown as a function of the fitting range $x$ 
(see text). Filled circles in (a) are experimental data from
Ref.~\cite{rtokuokh}. Dashed lines mark the result of the BEC description.}
\end{figure}

The dependence on window size $x_1$ of the exponent $\alpha_1$
is shown in Fig.~\ref{alphas}(a), along with data extracted from 
experimental reports on $\rm TlCuCl_3$ \cite{rtokuokh}. A first 
unambiguous conclusion from Fig.~\ref{alphas} is that the fitted 
exponents become smaller as the window size is decreased. In particular,
a large window leads to overestimation of $\alpha_1$, as is shown by 
performing the same varying-window fit to the experimental data for 
$T_c (h)$ [Fig.~\ref{alphas}(a)]. It is also clear that the 
extrapolated values approach the result obtained from the BEC 
description, $\alpha_{\rm BEC} = 1.5$. Higher exponents suggested 
by previous results are seen to be a product of the fitting procedure 
employed, and are in fact consistent with the BEC scenario at the 
transition [Fig.~\ref{alphas}(a)]. Most remarkably, the lines for all 
interdimer couplings not only scale to the same limit but do so with 
the same slope for all $x$, demonstrating that the 
critical behavior of the system is truly universal for all ratios of 
bandwidth to spin gap. 

Results for $\alpha_2$ are shown in Fig.~\ref{alphas}(b), and are again 
consistent with the BEC description as $x_2 \rightarrow 0$. Here the 
spread of data points is smaller than for $\alpha_1(x_1)$, and the slope 
of $\alpha_2(x_2)$ is less steep. These results may be attributed to the 
less significant role of quantum fluctuations near $h_{c2}$, which leads 
to smaller deviations from a mean-field scaling regime. 

It must be emphasized that the universal scaling behavior is obtained
only for very narrow windows, meaning very low temperatures. Indeed, 
from Fig.~\ref{alphas} we are unable to determine a temperature below 
which true scaling emerges, and our results indicate that the BEC regime 
is approached only as $T \rightarrow 0$. At both critical fields, the 
physical origin of departure from universal scaling may be found in the 
approximations inherent to the BEC description \cite{bec}. The most 
important of these are temperature-driven renormalization of the 
quasiparticle effective mass ({\it i.e.}~reduction in curvature at 
the band minimum) and of the effective chemical potential 
({\it i.e.}~band narrowing with concomitant increase in spin gap).

In summary, we have used an unbiased numerical technique to study 
field-induced magnetic ordering transitions in cubic systems of 
antiferromagnetically coupled dimers. The phase diagram contains 
two quantum critical lines, $h_{c1}(J'/J)$ and $h_{c2}(J'/J)$. Both 
transitions are continuous, as is the variation in physical properties 
of the intermediate AF phase, as a function of applied magnetic field. 
The critical scaling exponents $\alpha$ extracted for the transition 
temperature close to $h_{c1}$ and $h_{c2}$ are found to be independent 
of the interdimer coupling ratio, demonstrating that, despite large 
quantitative differences in their magnon dispersion relations, the 
physical properties of the coupled-dimer systems KCuCl$_3$ and TlCuCl$_3$ 
display the same universal scaling behavior. The values of $\alpha$ 
are consistent with a description of the ordering transitions as the 
Bose-Einstein condensation of magnon excitations, $\alpha_{\rm BEC} = 
1.5$. However, this regime is obtained only in the limit of temperatures 
very small compared to the interdimer coupling, and thus is difficult to 
access experimentally. Finally, the fits to the exponents were shown to 
depend sensitively on the window size, with the consequence that $\alpha$ 
is easily overestimated. This result offers an explanation for the wide 
range of scaling exponents reported in the literature.

\par
We thank F. Alet, A. Honecker, A. L\"auchli, M. Matsumoto, T. Roscilde, 
A. Sandvik, M. Sigrist, H. Tanaka, and M. Troyer for helpful discussions. 
We acknowledge financial support from the NSF, under Grant No. 
DMR-0089882, and from the Swiss NSF. Computational support was provided 
by the USC Center for High Performance Computing and Communications, by 
the NERSCC, and by Asgard at ETH Z\"urich.


\begin{thebibliography}{99}
\bibitem{runs} We have performed simulations for system sizes $L \le 
20$. However, data obtained for $L = 20$ neither improves the results 
nor reduces the error bars appreciably, and we have restricted production 
runs to system sizes $L \le 16$.
\bibitem{errors} The errors $\Delta \alpha_1$ in
Fig.~\ref{alphas}(a) are due to uncertainties in $h_{c1}$, which 
demand a two parameter fit. Because $h_{c2}$ is known, $\alpha_2$
can be obtained from a single parameter fit, resulting in
correspondingly smaller uncertainties $\Delta \alpha_2$, which are smaller than
the symbols of
Fig.~\ref{alphas}(b). The 
error quoted for the measured critical field~\cite{rtokuokh}, $H_{c1} 
= 5.7(1)$T, yields a 25\% error in $\alpha_1$, which is not shown 
in Fig.~\ref{alphas}.
\end{thebibliography}
\end{document}